\documentclass[onecolumn,showpacs,10pt]{article}

\def\mytitle#1{\setcounter{equation}{0}
\setcounter{footnote}{0}
\begin{flushleft}\Large\textbf{#1}\end{flushleft}
\vspace{0.25cm}}
\def\myname#1{\leftline{{\large #1}}\vspace{-0.13cm}}
\def\myplace#1#2{\small\begin{flushleft}\textit{#1}\\
\texttt{#2}\end{flushleft}}

\usepackage{graphicx}% Include figure files
\usepackage{amsmath, amssymb, graphics}
\newcommand{\mathsym}[1]{{}}

\begin{document}

\mytitle{\Large{Study of Some Cosmological Parameters for
Interacting New Holographic Dark Energy Model in \textit{f(T)}
Gravity}}

\vskip0.2cm \myname{Chayan
Ranjit\footnote{chayanranjit@gmail.com}} \myplace{Department of
Mathematics, Egra S. S. B. College, Purba Medinipur-721429, W.B.
India.}{}\vskip0.2cm

\vskip0.2cm \myname{ Prabir Rudra\footnote{prudra.math@gmail.com}}
\myplace{Department of Mathematics, Asutosh College, Kolkata-700
026, India.}{}\vskip0.2cm

\date{\today}

\begin{abstract}
The present work is based on the idea of an interacting framework
of new holographic dark energy with cold dark matter in the
background of $f(T)$ gravity. Here, we have considered the flat
modified Friedmann universe for $f(T)$ gravity which is filled
with new Holographic dark energy and dark matter. We have derived
some cosmological parameters like Deceleration parameter, EoS
parameter, State-finder parameters, Cosmographic parameters, {\it
Om} parameter and graphically investigated the nature of these
parameters for the above mentioned interacting scenario. The
results are found to be consistent with the accelerating universe.
Also we have graphically investigated the trajectories in $\omega
$--$ \omega'$ plane for different values of the interacting
parameter and explored the freezing region and thawing region in
$\omega $--$ \omega'$ plane. Finally, we have analyzed the
stability of this model.
\end{abstract}

\section{\normalsize\bf{Introduction}} Accelerated phenomenon of
the universe is now well established observable fact but the
physical origin of this acceleration is unknown till now. There
are so many observational studies like the different luminosity
type Ia supernovae (SNIa) \cite{Gold2009,
Nolta2009,Bachall1999,Perlmutter1999}, in associated with Large
scale Structure \cite{Abazajian2004,Abazajian2005} and Cosmic
Microwave Background (CMB) \cite{Perlmutter1998,Riess,Riess1,
Bennet,Sperge} radiation which strongly support that our universe
is currently expanding with an acceleration. Many candidates are
considered as the main responsible for this observational facts of
expanding scenario which triggered out a new type of matter which
violates the strong energy condition i.e., $\rho+p<0$. The energy
density of such mysterious matters which have negative pressure
thats leads to cosmic acceleration, are known as Dark Energy (DE)
\cite{Sahni2006,Sahni2006b,Seikel,Clarkson,Liu}. Recent
observational studies indicate that Dark energy occupies about
$73\%$ of the energy of our universe, while dark matter
contribution is $\sim$$23\%$ and the usual baryonic matter is
$\sim$$4\%$. Dark energies are separated depending on their
equation of state (EOS). In the case of quintessence type DE, the
EOS is $-1 <\omega<1/3$, whereas for phantom type DE, EOS is $
\omega<-1$ but in the case of quintom type DE, the EOS can cross
the phantom divide $ \omega=-1$ from both sides. The most charming
and simplest candidate for DE is the cosmological constant
$\Lambda$. However, it is known to all that the cosmological
constant suffers two serious theoretical problems, i.e. the
cosmological constant problem and the coincidence problem. Hence,
for seeking the most reliable model for describing the DE
phenomenon, different methods have been adopted and for that
different dynamical DE models have been
developed.\\

There are another point of view to describe that acceleration
scenario. Recently many researchers consider Modified Gravity as
the alternating candidate of dark energy in support of
accelerating phenomenon. The basic idea of Modified Gravity is
that a concept of modification of gravitational theory and as a
result it provides a very natural gravitational alternative for
exotic matter. Without commencing negative kinetic term of dark
energies, Modify Gravity models can explain the phantom or
non-phantom or quintom phase of the universe, not only that, it
also illustrates about the early time inflation then transition
from deceleration to late time acceleration. The most popular
String/M-theory also supports Modify Gravity models. In recent day
scenario, it is treated as the rival of General Relativity.
Although there are some rigid constraints in modify gravity
theory, some particular form of modified gravity obtains a great
attention of current researchers. Ricci curvature $R$ of
Lagrangian is one of them. Different form of Ricci curvature have
been considered \cite{Vollick,Briscese, Carroll,Abdalla,Linder} to
explain the accelerating phase of the Universe. In this regards,
some remarkable work was done by Nojiri et al
\cite{Nojiri2003,Nojiri2007} where they considered the modified
Lagrangian as $R+ R^m + R^{-m}$ and as a result they obtained an
inflation at an early stage and also a late time of accelerated
expansion. These result lead to open new window of gravity
dependent modern research. The most reliable modified gravity
models incorporate $f(R)$ gravity (where $R$ represent the Ricci
Scalar Curvature) \cite{Nojiri2007a,Li,Nojiri2006}, $f(T)$ gravity
(where $T$ is the torsion scalar) \cite{Li2,Myrzakulov,Wu}, $f(G)$
gravity (where $G$ represents the Gauss-Bonnet invariant)
\cite{Rastkar,Nojiri2005}, $f(R,T)$ gravity \cite{Harko2011,
Jamil2012}, $f(R,G)$ gravity \cite{Bamba2010,Myrzakulov2013} and
so on. Recently, reconstruction between different dark energy
models become a very appealing scenario in modern cosmological
study. In 2008, Setare et al. \cite{Setare2008} studied the
cosmological proposition of the correspondence of Holographic dark
energy model and Gauss-Bonnet dark energy model. That showed the
way to explanation of the accelerated expansion of the universe by
imposing specific constraints. In \cite{Liu2009} Liu et al. talked
about the New Agegraphic Dark Energy (NADE) model in the framework
of the Brans-Dicke theory with the help of EoS and deceleration
parameters and finally they showed that the EoS parameter has a
quintom-like behavior for that model and it indicates the
accelerated expansion of the universe. The NADE in $f(R)$ gravity
model has been discussed in \cite{Setare2010} and as a result it
found that there may exist a phantom-like universe. Also in
\cite{Jamil2010} the NADE model is correspond with Horava-Lifshitz
gravity which specified that the accelerated expanding universe is
consistent with cosmological observations. The reconstruction of
Entropy-Corrected Holographic Dark Energy (ECHDE) in the $f(G)$
gravity also been investigated to explain the expanding universe
with acceleration in \cite{Setare2010b}. Recently the HDE model in
the framework of the $f(G)$ gravity has been discussed in
\cite{Jawad2013} and the different phenomenon for the accelerating
universe are explained.\\

The main purpose of this work is to analyzed some cosmological
parameters for interacting new Holographic dark energy model in
$f(T)$ gravity model. For this reason, homogeneous, isotropic
modified FRW for $f(T)$ gravity has been considered and after that
conservation equations for interacting new Holographic dark energy
and dark matter have been solved in Section 2. Then density and
pressure are converted into dimensionless quantity. After that
Deceleration parameter, EoS parameter, $\omega $--$ \omega'$ plane
analysis, State-finder parameters, Cosmographic parameters and
{\it Om} parameter are converted in terms of current density,
pressure and Hubble expansion rate parameter and graphically
analyzed in Section 3 respectively. In Section 4, stability of
this model have been analyzed. Finally, the results of the paper
are summarized in Section 5.

\section{\bf{Field equations and their solutions}}

The metric of a spatially flat homogeneous and isotropic universe
in FRW model is given by
\begin{equation}
ds^{2}=dt^{2}-a^{2}(t)[dr^{2}+r^{2}(d\theta^{2}+sin^{2}\theta
d\phi^{2})]
\end{equation}
where $a(t)$ is the scale factor.

The Einstein field equation are given by
\begin{equation}
H^{2}=\frac{1}{3}\rho
\end{equation}
and
\begin{equation}
2\dot{H}+3H^{2}=-p
\end{equation}
where $\rho$ and $p$ are energy density and isotropic pressure
respectively (choosing $8\pi G=c=1$)and $H=\frac{\dot{a}}{a}$ is
the Hubble parameter.

Let us we start with $f(T)$ gravity models where the action is
given by
\begin{equation}
S=\frac{1}{2\kappa^{2}}\int dx^{4}[\sqrt{-g}f(T)+L_{m}]
\end{equation}
where $T$ is the torsion scalar, $f(T)$ is general differentiable
function of the torsion and $L_{m}$ corresponds to the matter
Lagrangian, $\kappa^{2}=8\pi G=1$. Therefore the field equations
for modified FRW model are given by
\begin{equation}\label{2.5}
3H^{2}=\rho_{MG}=\rho+\rho_{T}
\end{equation}
and
\begin{equation}\label{2.6}
2\dot{H}+3H^{2}=-p_{MG}=-(p+p_{T})
\end{equation}
where
\begin{equation}
\rho_{T}=\frac{1}{2}(2Tf_{T}-f+6H^{2}),~~~p_{T}=-\frac{1}{2}[-8\dot{H}Tf_{TT}+(2T-4\dot{H})f_{T}-f+4\dot{H}+6H^{2}]
\end{equation}
and $T=-6H^{2}$.

We consider the power law model $f(T)=T^{n}$, where $n$ is a
constant. Restricting ourselves to $n=2$ for simplicity, we obtain
\begin{equation}\label{2.8}
\rho_{T}=3H^{2}(18H^{2}+1),~~~~p_{T}=-2\dot{H}-3H^{2}(24\dot{H}+18H^{2}+1)
\end{equation}

 Also we consider that our universe is filled with new Holographic
dark energy and dark matter. So we assume,
$\rho=\rho_{x}+\rho_{m}$ and $p=p_{x}+p_{m}$. Here $\rho_{x}$,
$p_{x}$ and $\rho_{m}$, $p_{m}$ are respectively the energy
density and pressure for new holographic dark energy and dark
matter. Although the dark matter has negligible pressure, i.e.,
$p_{m}\approx 0$ but here we taken into account the non-zero value
of $p_{m}$(which is very small value).

The energy conservation equation is given by
\begin{equation}\label{2.9}
\dot{\rho}+3\frac{\dot{a}}{a}(\rho+p)=0
\end{equation}
Now we consider the model of interaction between dark matter and
the new holographic dark energy model, through a phenomenological
interaction term Q. Keeping into consideration the fact that the
Supernovae and CMB data determine that decay rate should be
proportional to the present value of the Hubble parameter. The
interaction term describes the energy flow between the two fluids.
Therefore the conservation equation (\ref{2.9}) becomes
\begin{equation}\label{2.10}
\dot{\rho}_{x}+3H(\rho_{x}+p_{x})=-Q
\end{equation}
and
\begin{equation}\label{2.11}
\dot{\rho}_{m}+3H(\rho_{m}+p_{m})=Q
\end{equation}
There are many different forms of Q in the literature, we choose
the following form of Q as
\begin{equation}
Q=3\delta H\rho_{m}
\end{equation}
where $\delta$ (may positive or negative)is the interaction
parameter. In our present study we take $\delta\geq 0$.

 The new Holographic dark energy density is given by
\begin{equation}\label{2.13}
\rho_{x}=3(\alpha H^{2}+\beta \dot{H})
\end{equation}
where $\alpha$ and $\beta$ are constants.
 Now the EOS of dark matter is given by
$p_{m}=\omega_{m}\rho_{m}$, where $\omega_{m}$ is very small and
so the equation (\ref{2.11}) becomes
\begin{equation}
\rho_{m}=\rho_{m0}a^{-3(1+\omega_{m}-\delta)}
\end{equation}
 and consequently from the equation (\ref{2.5}), we have
\begin{equation}
H^{2}=\frac{\sqrt{-\frac{\beta}{a^{3(1+\omega_{m})}}}\left[\alpha
\sqrt{-\frac{a^{3(1+\omega_{m})}}{\beta}}
J_{-U}(V)C+\sqrt{-\frac{6a^{3\delta}\rho_{m0}}{\beta}}\left(2J_{-1+U}(V)+\left(J_{-1-U}(V)-
J_{1-U}(V)\right)C\right)\right]}{36\left(J_{U}(V)+J_{-U}(V)C\right)}
\end{equation}
where $J_{U}(V)$ is a Bessel function of first kind,
$U=\frac{2\alpha}{3\beta(1-\delta+\omega_{m})}$ and
$V=-\frac{\sqrt{\frac{32}{3}\rho_{m0}}
a^{-\frac{3}{2}(1-\delta+\omega_{m})}}{\beta(1-\delta+\omega_{m})}$
and $\rho_{m0}$ is the present value of matter density (at $a=1$)
and $C$ is the arbitrary integration constant.

Now we define:
\begin{equation}\label{2.16}
\begin{array}{l}
\widetilde{H}^{2}=\frac{H^{2}}{H_{0}^{2}};~~~~~~\\\\
\widetilde{\rho}_{m}=\frac{\rho_{m}}{3H_{0}^{2}};~~~~
\widetilde{\rho}_{x}=\frac{\rho_{x}}{3H_{0}^{2}};~~~~
\widetilde{\rho}_{T}=\frac{\rho_{T}}{3H_{0}^{2}};~~~~~~\\\\
\widetilde{p}_{m}=\frac{p_{m}}{3H_{0}^{2}};~~~~
\widetilde{p}_{x}=\frac{p_{x}}{3H_{0}^{2}};~~~~
\widetilde{p}_{T}=\frac{p_{T}}{3H_{0}^{2}};~~~~~~\\\\
\widetilde{\Omega}_{m0}=\frac{\rho_{m0}}{3H_{0}^{2}};~~~~
\widetilde{\Omega}_{x0}=\frac{\rho_{x0}}{3H_{0}^{2}};~~~~
\widetilde{\Omega}_{T0}=\frac{\rho_{T0}}{3H_{0}^{2}};~~~~
\end{array}
\end{equation}

where $H_{0}$ is the present value of the Hubble parameter,
$\widetilde{H}$ is the Hubble expansion rate, current density
parameters are $\Omega_{m0}$ and $\Omega_{x0}$ for matter and dark
energy respectively. The scale factor $a$ is expressed in terms of
redshift z as
\begin{equation}
a=\frac{1}{1+z}
\end{equation}
Therefore from the equations
(\ref{2.5})-(\ref{2.8}),(\ref{2.10})-(\ref{2.16}) we have

\begin{equation}
\widetilde{\rho}_{m}=\widetilde{\Omega}_{m0}(1+z)^{3(1-\delta+\omega_{m})}
\end{equation}

\begin{equation}
\widetilde{\rho}_{T}=(18H^{2}+1)\widetilde{H}^{2}
\end{equation}

\begin{equation}
\widetilde{\rho}_{x}=\alpha \widetilde{H}^{2}-
\frac{\beta(1+z)}{2}\frac{d\widetilde{H}^{2} }{dz}
\end{equation}

\begin{equation}
\widetilde{p}_{m}=\omega_{m}\widetilde{\Omega}_{m0}(1+z)^{3(1-\delta+\omega_{m})}
\end{equation}

\begin{equation}
\widetilde{p}_{T}=\frac{1}{3\beta}\left[\left((2\alpha-3\beta)
+18H^{2}(4\alpha-3\beta)
\right)\widetilde{H}^{2}-2(36H^{2}+1)\widetilde{\rho}_{x}\right]
\end{equation}
and
\begin{equation}
\widetilde{p}_{x}=-\delta\widetilde{\rho}_{m}-\alpha
\widetilde{H}^{2}+\frac{1}{3}(\alpha+\beta)(1+z)\frac{d\widetilde{H}^{2}
}{dz}-\frac{1}{6}\beta(1+z)^{2}\frac{d^2\widetilde{H}^{2} }{dz^2}
\end{equation}
where we use the relations
$\frac{\dot{H}}{H^{2}}=-\frac{(1+z)}{2\widetilde{H}^{2}}
\frac{d\widetilde{H}^{2} }{dz}$ \&
$\frac{\ddot{H}}{H^{3}}=\frac{(1+z)}{2\widetilde{H}^{2}}
\frac{d\widetilde{H}^{2}
}{dz}+\frac{(1+z)^{2}}{2\widetilde{H}^{2}}\frac{d^{2}\widetilde{H}^{2}
}{dz^{2}}$.

\section{{\normalsize\bf{Some Well-Known Cosmological
Parameters}}}
\subsection{\normalsize\bf{Deceleration Parameter}}

Deceleration parameter plays a very important role for any
cosmological model to determine the cosmic acceleration. A
negative value of q represents cosmic acceleration, whereas a
positive q gives a decelerating universe. The deceleration
parameter is given by
\begin{equation}
q=-1-\frac{\dot{H}}{H^{2}}=-1+\frac{(1+z)}{2\widetilde{H}^{2}}
\frac{d\widetilde{H}^{2} }{dz}
\end{equation}
and in this present scenario, the expression for $q$ is obtained
as
\begin{equation}
\begin{array}{ll}
q=-1+\left(18 \left(\frac{1}{1+z}\right)^{-1+\frac{3}{2}
\left(1+\omega _m\right)} \left(J_{U}(V)+J_{-U}(V) C\right) H_0^2
\right. \left(\left(\left(\frac{1}{1+z}\right)^{1-\frac{3}{2}
\left(1+\omega _m\right)} \sqrt{-\frac{1}{\beta }} \beta
\left(\left(\frac{1}{1+z}\right)^{\frac{3}{2} \left(1+\omega
_m\right)} \alpha  \sqrt{-\frac{1}{\beta }}
J_{-U}(V) C+\right.\right.\right.\\
\left.\left.\left.\sqrt{6} \left(\frac{1}{1+z}\right)^{3 \delta
/2} \left(2 J_{-1+U}(V)+(J_{-1-U}(V)-J_{1-U}(V)) C\right)
\sqrt{-\frac{\rho_{m0}}{\beta }}\right) \left(1+\omega
_m\right)\right)\right/\left(24 \left(J_U(V)+J_{-U}(V) C\right)
H_0^2\right)+\\ \left(\left(\frac{1}{1+z}\right)^{-\frac{3}{2}
\left(1+\omega _m\right)} \sqrt{-\frac{1}{\beta }} \beta  \right.
\left(-3 \sqrt{\frac{3}{2}} \left(\frac{1}{1+z}\right)^{1+\frac{3
\delta }{2}} \delta  \left(2 J_{-1+U}(V)+(J_{-1-U}(V)-J_{1-U}(V))
C\right) \sqrt{-\frac{\rho _{m0}}{\beta }}-\right.\\
\left.\left.\left.\frac{3}{2}
\left(\frac{1}{1+z}\right)^{1+\frac{3}{2} \left(1+\omega
_m\right)} \alpha  \sqrt{-\frac{1}{\beta }} J_{-U}(V) C
\left(1+\omega _m\right)\right)\right)\right/
\left.\left.\left.\left(36 \left(J_U(V)+J_{-U}(V) C\right)
H_0^2\right)\right)\right)\right/\\
\left(\sqrt{-\frac{1}{\beta }} \beta  \right.
\left(\left(\frac{1}{1+z}\right)^{\frac{3}{2} \left(1+\omega
_m\right)} \alpha  \sqrt{-\frac{1}{\beta }} J_{-U}(V) C+\right.
\left.\left.\sqrt{6} \left(\frac{1}{1+z}\right)^{3 \delta /2}
\left(2 J_{-1+U}(V)+(J_{-1-U}(V)-J_{1-U}(V)) C\right)
\sqrt{-\frac{\rho _{m0}}{\beta }}\right)\right)
\end{array}
\end{equation}
The deceleration parameter is plotted against the redshift
parameter in Fig 1.
\begin{figure}
~~~~~~~~~~~~~~~~~~~~~~~~~~~\includegraphics[scale=0.85]{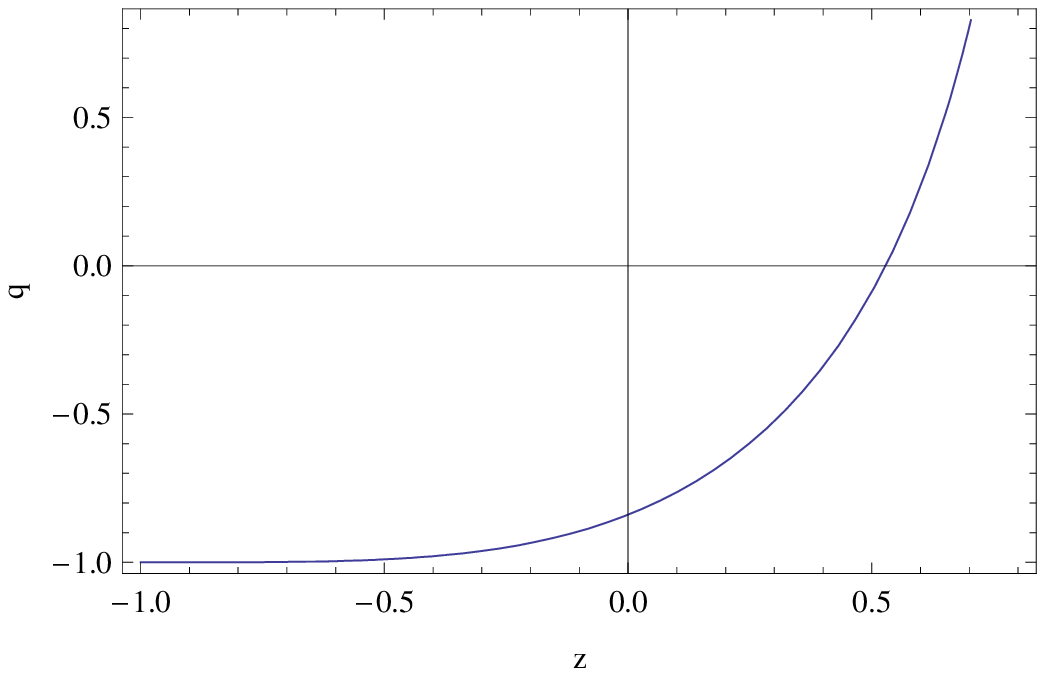}\\\label{4.1}
\vspace{2mm}
~~~~~~~~~~~~~~~~~~~~~~~~~~~~~~~~~~~~~~~~~~~~~~~~~~~~~~~~Fig.1~~\\
\vspace{4mm} Fig.1 shows the variation of deceleration parameter
(q) against the redshift (z). \vspace{6mm}
\end{figure}
\subsection{\normalsize\bf{EoS Parameter}}
The nature of matter content of the universe is depicted by the
Equation of State (EoS) parameter $\omega$. Basically its gives an
idea about the era of the universe. When $\omega=0,1/3$ and $1$,
it can be predicted that the universe is in the dust era, the
radiation era and stiff fluid respectively whereas $\omega=-1/3,
-1$ and $\omega<-1$ represent the quintessence DE, $\Lambda$CDM
era and Phantom era respectively. The EoS parameter can be
obtained as

\begin{equation}
\omega=\frac{2q-1}{3}
\end{equation}

\begin{figure}
~~~~~~~~~~~~~~~~~~~~~~~~~~~~\includegraphics[scale=0.85]{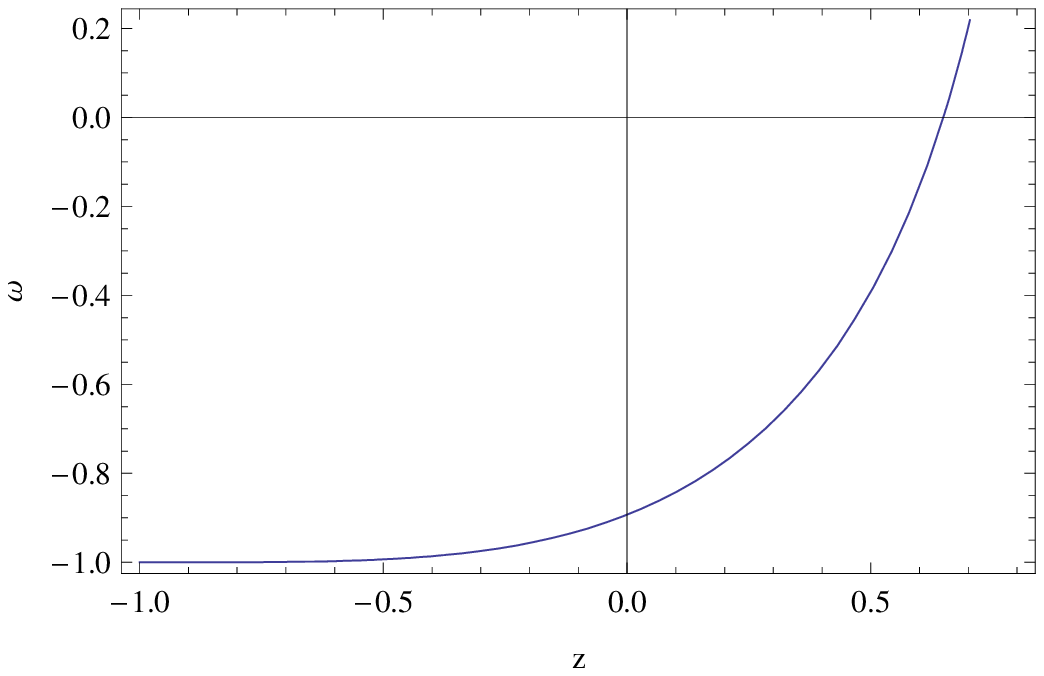}\\
\vspace{2mm}
~~~~~~~~~~~~~~~~~~~~~~~~~~~~~~~~~~~~~~~~~~~~~~~~~~~~Fig.2~~\\
\vspace{4mm} Fig.2 depicts the change of EoS parameter against the
redshift. \vspace{6mm}
\end{figure}
Here we have obtained it for the given model as:
\begin{equation}
\begin{array}{ll}\label{4.2}
\omega=-\frac{1}{3}\\+\frac{2}{3}\left[-1+\left(18
\left(\frac{1}{1+z}\right)^{-1+\frac{3}{2} \left(1+\omega
_m\right)} \left(J_{U}(V)+J_{-U}(V) C\right) H_0^2 \right.
\left(\left(\left(\frac{1}{1+z}\right)^{1-\frac{3}{2}
\left(1+\omega _m\right)} \sqrt{-\frac{1}{\beta }} \beta
\left(\left(\frac{1}{1+z}\right)^{\frac{3}{2} \left(1+\omega
_m\right)} \alpha  \sqrt{-\frac{1}{\beta }}
J_{-U}(V) C+\right.\right.\right.\right.\\
\left.\left.\left.\left.\sqrt{6} \left(\frac{1}{1+z}\right)^{3
\delta /2} \left(2 J_{-1+U}(V)+(J_{-1-U}(V)-J_{1-U}(V)) C\right)
\sqrt{-\frac{\rho_{m0}}{\beta }}\right) \left(1+\omega
_m\right)\right)\right/\left(24 \left(J_U(V)+J_{-U}(V) C\right)
H_0^2\right)+\right.\\
\left.\left(\left(\frac{1}{1+z}\right)^{-\frac{3}{2}
\left(1+\omega _m\right)} \sqrt{-\frac{1}{\beta }} \beta  \right.
\left(-3 \sqrt{\frac{3}{2}} \left(\frac{1}{1+z}\right)^{1+\frac{3
\delta }{2}} \delta  \left(2 J_{-1+U}(V)+(J_{-1-U}(V)-J_{1-U}(V))
C\right) \sqrt{-\frac{\rho _{m0}}{\beta }}-\right.\right.\\
\left.\left.\left.\left.\frac{3}{2}
\left(\frac{1}{1+z}\right)^{1+\frac{3}{2} \left(1+\omega
_m\right)} \alpha  \sqrt{-\frac{1}{\beta }} J_{-U}(V) C
\left(1+\omega _m\right)\right)\right)\right/
\left.\left.\left.\left(36 \left(J_U(V)+J_{-U}(V) C\right)
H_0^2\right)\right)\right)\right/\right.\\
\left.\left(\sqrt{-\frac{1}{\beta }} \beta  \right.
\left(\left(\frac{1}{1+z}\right)^{\frac{3}{2} \left(1+\omega
_m\right)} \alpha  \sqrt{-\frac{1}{\beta }} J_{-U}(V) C+\right.
\left.\left.\sqrt{6} \left(\frac{1}{1+z}\right)^{3 \delta /2}
\left(2 J_{-1+U}(V)+(J_{-1-U}(V)-J_{1-U}(V)) C\right)
\sqrt{-\frac{\rho _{m0}}{\beta }}\right)\right)\right]
\end{array}
\end{equation}
The plot of the EoS parameter against the redshift parameter is
obtained in Fig 2.

\subsection{\normalsize\bf{\Large{$\omega$--$ \omega'$} Plane
Analysis}}

The $\omega$--$ \omega'$(where prime denotes derivatives with
respect to $x = ln~a$) plane analysis is very significant tool in
our modern days cosmological analysis. Basically, it has been used
to distinguish different DE models through trajectories on its
plane. At first this approach was proposed by Caldwell and Linder
in 2005 \cite{Caldwell2005}. Initially, this method has been
applied on quintessence DE model which leads to two classes of its
plane, one of which has been known as thawing region where the
area occupied by the region as $\omega>0$ for $\omega'<0$ on
$\omega$--$ \omega'$ plane while the other known as freezing
region where $\omega<0$ for $\omega'<0$ and it is also notable
that the expansion of the universe is comparatively more
accelerating in freezing region. Recently, this tool has been
widely applied to other well-known dynamical DE models such as
more general form of quintessence \cite{Scherrer}, Pilgrim DE
\cite{Sharif2013a,Sharif2013b,Sharif2014}, quintom \cite{Guo},
phantom \cite{Chiba}, polytropic DE \cite{Malekjani} and others.
Here we try to analyze and emphasize such regions with the help of
this method. Now $\omega'$ can be obtained by differentiating Eq.
(\ref{4.2}) with respect to $x$. The plot for $\omega$ versus $
\omega'$ for our predicted model with different value of
$\delta=1, 0, 0.5$ are given in Figs. 3-5 respectively. Also it
can be observed from those figures that the freezing regions are
possible for non-interacting scenario i.e., $\delta=0$ and for
interacting scenario with $\delta=0.5$, whereas for high
interaction scenario, $\delta=1.0$ the thawing region is a
possibility with the assumption that the other parameters are
fixed at same value as the previous cases. Again it can be
observed that $\omega'\rightarrow 0$ when $\omega\rightarrow -1$
i.e., $\Lambda$CDM limit may be possible only the in case when $\delta=0$.\\

\begin{figure}
\includegraphics[scale=0.75]{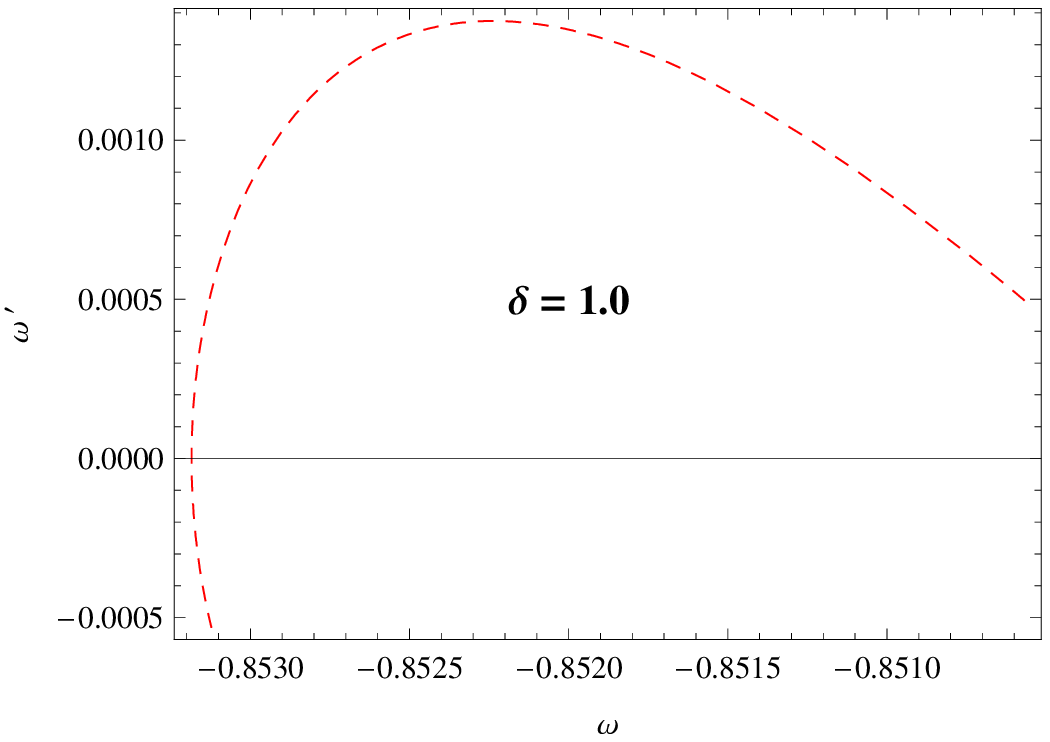}~~
\includegraphics[scale=0.75]{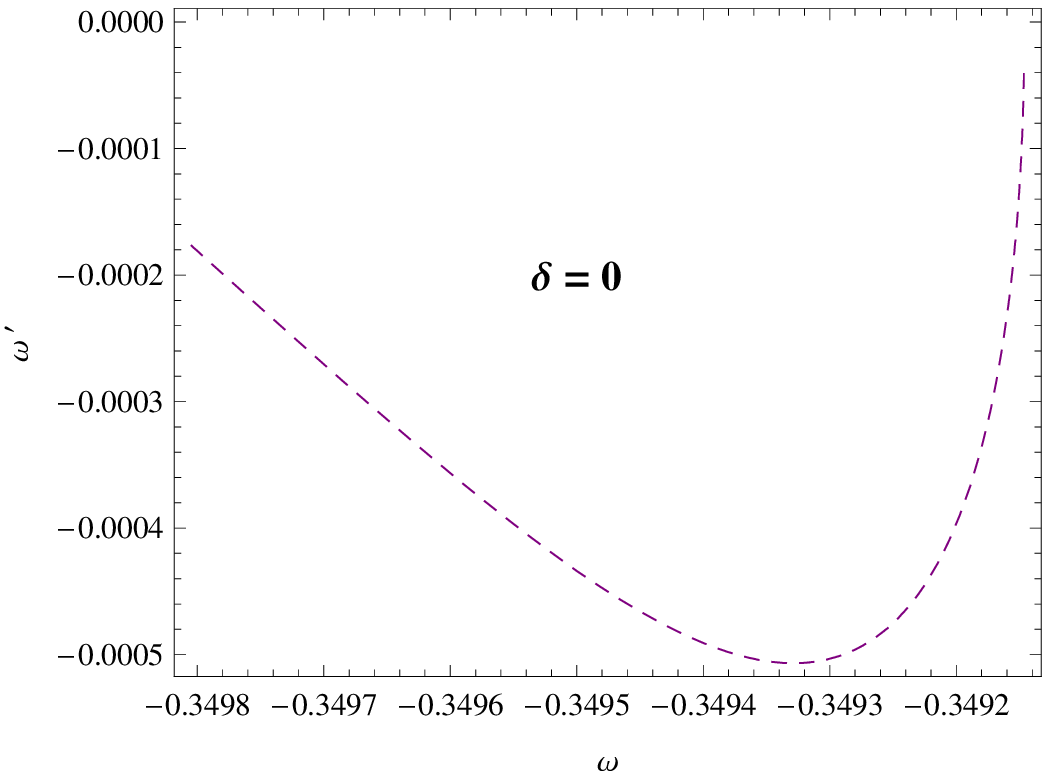}~~\\
\vspace{1mm} ~~~~~~~~~~~~~~~~~~~~~~~~~~~~~~~~~~Fig.3
~~~~~~~~~~~~~~~~~~~~~~~~~~~~~~~~~~~~~~~~~~~~~~~~~~~~~~~~~~~~~~~Fig.4~~\\\\
\vspace{1mm}~~~~~~~~~~~~~~~~~~~~~~~~~~~~
\includegraphics[scale=0.85]{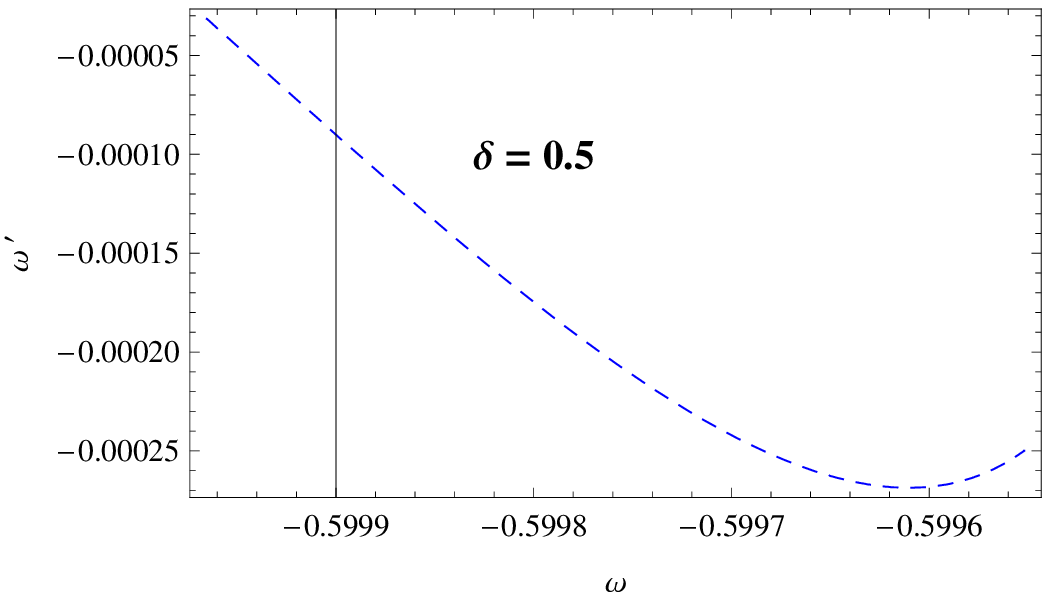}~~\\
\vspace{1mm}
~~~~~~~~~~~~~~~~~~~~~~~~~~~~~~~~~~~~~~~~~~~~~~~~~~~~~~~~~~~~~~~~~~~Fig.5~~\\
\vspace{1mm} Fig. 3-5 show the plot of $\omega$--$\omega'$ for
$\delta=1.0,0,0.5$ respectively with other parameters fixed at
$C=0.005$, $\omega_{m}=0.3$, $\rho_{m0}=0.01$, $\alpha=0.95$,
$\beta=-2.1$ and $H_{0}=73$.
\end{figure}

\subsection{\normalsize\bf{Statefinder Parameters}} Statefinder
Parameters solved the problem of discriminating between the
various candidates for dark energy model. In 2003 Sahni et
al\cite{Sahni2003a} proposed the trajectories in the \{$r,s$\}
plane corresponding to different cosmological models to depict
qualitatively different behavior. The statefinder diagnostic along
with future SNAP observations may perhaps be used to discriminate
between different dark energy models. The above statefinder
diagnostic pair are defined as follows:

\begin{equation}
r=1+3\frac{\dot{H}}{H^{2}}+\frac{\ddot{H}}{H^{3}} ~~\text{and} ~~
s=\frac{r-1}{3(q-\frac{1}{2})}
\end{equation}
where $q$ is the deceleration parameter defined by
$q=-1-\frac{\dot{H}}{H^{2}}$ and $H$ is the Hubble parameter.
 These parameters

Now r and s can be written in terms of Hubble parameter H in the
following forms:
\begin{equation}
r=1+3\frac{\dot{H}}{H^{2}}+\frac{\ddot{H}}{H^{3}}
\end{equation}
and
\begin{equation}
s=-\frac{3H\dot{H}+\ddot{H}}{3H(2\dot{H}+3H^{2})}
\end{equation}
Or, equivalently in terms of $\widetilde{H}$ as follows:

\begin{equation}
r=1-\frac{(1+z)}{\widetilde{H}^{2}}\frac{d\widetilde{H}^{2} }{dz}
+\frac{(1+z)^{2}}{2\widetilde{H}^{2}}\frac{d^{2}\widetilde{H}^{2}
}{dz^{2}}
\end{equation}

\begin{equation}
s=\frac{-\frac{(1+z)}{\widetilde{H}^{2}}\frac{d\widetilde{H}^{2}
}{dz}+\frac{(1+z)^{2}}{2\widetilde{H}^{2}}\frac{d^{2}\widetilde{H}^{2}
}{dz^{2}}}{3\left(\frac{(1+z)}{\widetilde{H}^{2}}\frac{d\widetilde{H}^{2}
}{dz}-3\right)}
\end{equation}
where the explicit form of $r$ is as
\begin{equation}
\begin{array}{ll}
r=1-\left(36 \left(\frac{1}{1+z}\right)^{-1+\frac{3}{2}
\left(1+\omega _m\right)} \left(J_{U}(V)+J_{-U}(V) C\right) H_0^2
\right. \left(\left(\left(\frac{1}{1+z}\right)^{1-\frac{3}{2}
\left(1+\omega _m\right)} \sqrt{-\frac{1}{\beta }} \beta
\left(\left(\frac{1}{1+z}\right)^{\frac{3}{2} \left(1+\omega
_m\right)} \alpha  \sqrt{-\frac{1}{\beta }} J_{-U}(V)
C+\right.\right.\right.\\
\left.\left.\left.\sqrt{6} \left(\frac{1}{1+z}\right)^{3 \delta
/2} \left(2 J_{-1+U}(V)+(J_{-1-U}(V)-J_{1-U}(V)) C\right)
\sqrt{-\frac{\rho_{m0}}{\beta }}\right) \left(1+\omega
_m\right)\right)\right/\left(24 \left(J_{U}(V)+J_{-U}(V)
C\right) H_0^2\right)+\\
\left(\left(\frac{1}{1+z}\right)^{-\frac{3}{2} \left(1+\omega
_m\right)} \sqrt{-\frac{1}{\beta }} \beta  \right. \left(-3
\sqrt{\frac{3}{2}} \left(\frac{1}{1+z}\right)^{1+\frac{3 \delta
}{2}} \delta  \left(2 J_{-1+U}(V)+(J_{-1-U}(V)-J_{1-U}(V))
C\right) \sqrt{-\frac{\rho_{m0}}{\beta }}-\right.\\
\left.\left.\left.\frac{3}{2}
\left(\frac{1}{1+z}\right)^{1+\frac{3}{2} \left(1+\omega
_m\right)} \alpha  \sqrt{-\frac{1}{\beta }} J_{-U}(V) C
\left(1+\omega _m\right)\right)\right)\right/
\left.\left.\left.\left(36 \left(J_{U}(V)+J_{-U}(V) C\right)
H_0^2\right)\right)\right)\right/
\left(\sqrt{-\frac{1}{\beta }} \beta  \right.\\
\left(\left(\frac{1}{1+z}\right)^{\frac{3}{2} \left(1+\omega
_m\right)} \alpha  \sqrt{-\frac{1}{\beta }} J_{-U}(V) C+\right.
\left.\left.\sqrt{6} \left(\frac{1}{1+z}\right)^{3 \delta /2}
\left(2 J_{-1+U}(V)+(J_{-1-U}(V)-J_{1-U}(V))
C\right) \sqrt{-\frac{\rho_{m0}}{\beta }}\right)\right)+\\
\left(18 \left(\frac{1}{1+z}\right)^{-2+\frac{3}{2} \left(1+\omega
_m\right)} \left(J_{U}(V)+J_{-U}(V) C\right) H_0^2 \right.
\left(-\left(\left(\frac{1}{1+z}\right)^{2-\frac{3}{2}
\left(1+\omega _m\right)} \sqrt{-\frac{1}{\beta }} \beta
\left(\left(\frac{1}{1+z}\right)^{\frac{3}{2} \left(1+\omega
_m\right)} \alpha  \sqrt{-\frac{1}{\beta }} J_{-U}(V)
C+\right.\right.\right.\\
\left.\sqrt{6} \left(\frac{1}{1+z}\right)^{3 \delta /2} \left(2
J_{-1+U}(V)+(J_{-1-U}(V)-J_{1-U}(V)) C\right)
\sqrt{-\frac{\rho_{m0}}{\beta }}\right) \left.\left(1+\omega
_m\right) \left(1-\frac{3}{2} \left(1+\omega
_m\right)\right)\right)/\left(24 \left(J_{U}(V)+J_{-U}(V) C\right)
H_0^2\right)\\
+\left(\left(\frac{1}{1+z}\right)^{1-\frac{3}{2} \left(1+\omega
_m\right)} \sqrt{-\frac{1}{\beta }} \beta  \left(1+\omega
_m\right) \right. \left(-3 \sqrt{\frac{3}{2}}
\left(\frac{1}{1+z}\right)^{1+\frac{3 \delta }{2}} \delta  \left(2
J_{-1+U}(V)+(J_{-1-U}(V)-J_{1-U}(V))
C\right) \sqrt{-\frac{\rho_{m0}}{\beta }}-\right.\\
\left.\left.\frac{3}{2} \left(\frac{1}{1+z}\right)^{1+\frac{3}{2}
\left(1+\omega _m\right)} \alpha  \sqrt{-\frac{1}{\beta }}
J_{-U}(V) C \left(1+\omega _m\right)\right)\right)/\left(12
\left(J_{U}(V)+J_{-U}(V) C\right) H_0^2\right)+
\left(\left(\frac{1}{1+z}\right)^{-\frac{3}{2} \left(1+\omega
_m\right)} \sqrt{-\frac{1}{\beta }} \beta  \right.\\
\left(3 \sqrt{\frac{3}{2}} \left(\frac{1}{1+z}\right)^{2+\frac{3
\delta }{2}} \delta \left(1+\frac{3 \delta }{2}\right) \left(2
J_{-1+U}(V)+(J_{-1-U}(V)-J_{1-U}(V))
C\right) \sqrt{-\frac{\rho_{m0}}{\beta }}+\right.\\
\left.\left.\left.\frac{3}{2}
\left(\frac{1}{1+z}\right)^{2+\frac{3}{2} \left(1+\omega
_m\right)} \alpha  \sqrt{-\frac{1}{\beta }} J_{-U}(V) C
\left(1+\omega _m\right) \left(1+\frac{3}{2} \left(1+\omega
_m\right)\right)\right)\right)\right/\left.\left.\left.\left(36
\left(J_{U}(V)+J_{-U}(V) C\right)
H_0^2\right)\right)\right)\right/\\
\left(\sqrt{-\frac{1}{\beta }} \beta  \right.
\left(\left(\frac{1}{1+z}\right)^{\frac{3}{2} \left(1+\omega
_m\right)} \alpha  \sqrt{-\frac{1}{\beta }} J_{-U}(V) C+\right.
\left.\left.\sqrt{6} \left(\frac{1}{1+z}\right)^{3 \delta /2}
\left(2 J_{-1+U}(V)+(J_{-1-U}(V)-J_{1-U}(V)) C\right)
\sqrt{-\frac{\rho_{m0}}{\beta }}\right)\right)
\end{array}
\end{equation}

\begin{figure}
\includegraphics[scale=0.75]{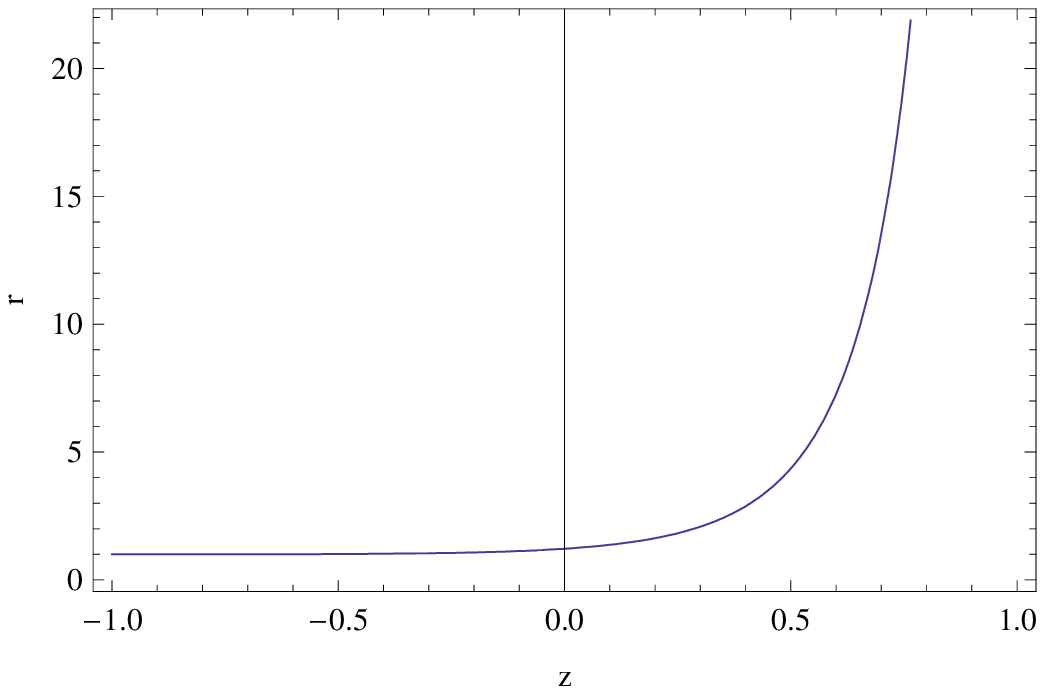}~~
\includegraphics[scale=0.75]{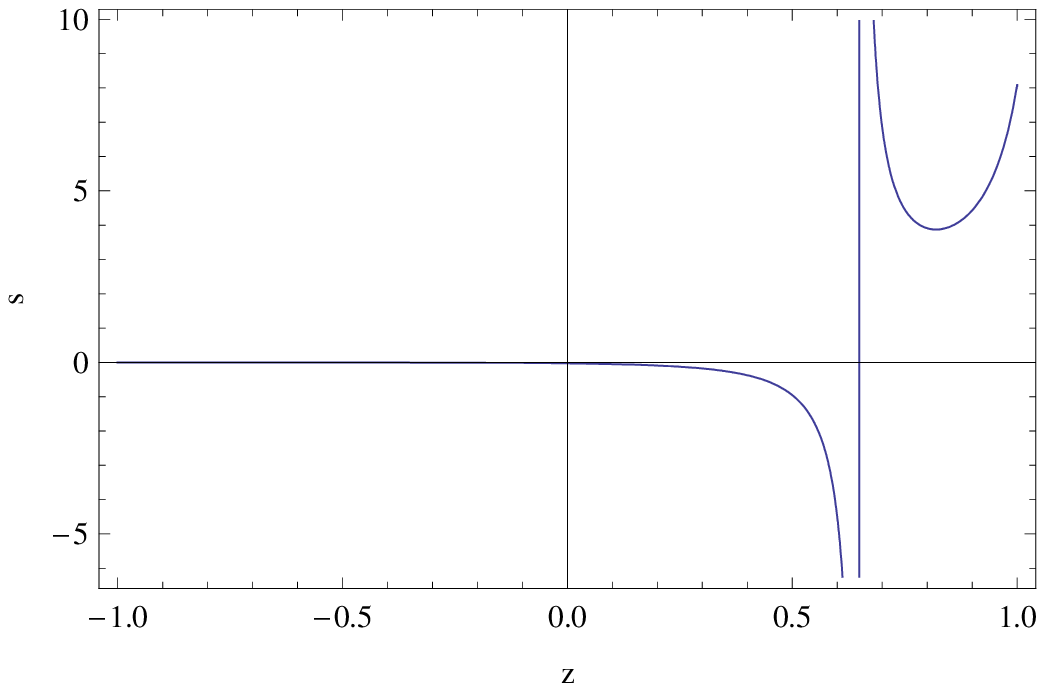}\\
\vspace{2mm}
~~~~~~~~~~~~~~~~~~~~~~~~~~~~~~~~Fig.6~~~~~~~~~~~~~~~~~~~~~~~~~~~~~~~~~~~~~~~~~~~~~~~~~~~~~~~~~~~~~~~~Fig.7\\
\vspace{6mm}~~~~~~~~~~~~~~~~~~~~~~~~~~~~~~~~~~~~
\includegraphics[scale=0.7]{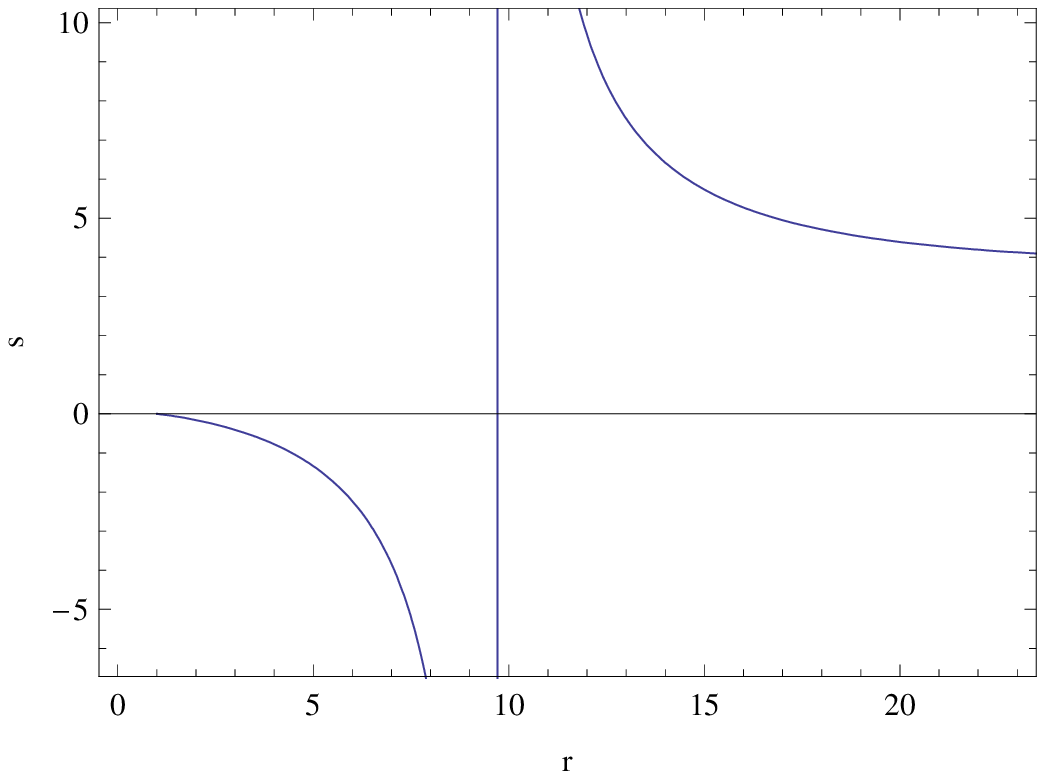}~~\\
\vspace{2mm}
~~~~~~~~~~~~~~~~~~~~~~~~~~~~~~~~~~~~~~~~~~~~~~~~~~~~~~~~~~~~~~~~Fig.8~~~~~~~~~~~~~~~~~~~~~~~~~~~~~~~~~~~~~~~~\\
\vspace{1mm} Fig. 6 shows the variation of the state statefinder
parameter r with redshift z. \\
Fig. 7 shows the variation of the state statefinder parameter s
with redshift z. \\
Fig. 8 shows the variation of the state
statefinder parameter s against r.\vspace{1mm}
\end{figure}

\subsection{\normalsize\bf{Cosmographic parameters}}

Standard candle is an astronomical ideal object that has a known
absolute magnitude. It is extremely important since by measuring
the apparent magnitude of the object, the distance can also be
determine. It helps to reconstruct the Hubble diagram, i.e, the
redshift - distance relation up to high redshift values and for
that it is commonly used to parameterized model where it can
verify the viability of the data and characterizing parameters.
There were some uncertainty as the method is model dependent.
Therefore some doubts come to mind when we examine the validity of
derived quantities depending on the present day values of the
deceleration parameters and the age of the universe. In order to
conquer such problem we need a model independent approach and for
that the alternative choice is cosmography where the scale factor
is expanded in Taylor series with respect to the cosmic time. Such
expansions not only show the way to distance-redshift relation, it
also supports the assumption of the Robertson-Walker metric. It
does not depend on the particular form of the solution of cosmic
equations. For that aim, the Cosmographic parameters that are
proportional to the coefficients of Taylor series expansion of the
scale factor with respect to the cosmic time defined as
\begin{equation}
q=-\frac{1}{aH^{2}}\frac{d^{2}a}{dt^{2}}=-\frac{1}{H}\frac{d}{da}(aH)=-1+\frac{(1+z)}{\widetilde{H}^{2}}
\frac{d\widetilde{H}^{2} }{dz}
\end{equation}
\begin{equation}
J=r=\frac{1}{aH^{3}}\frac{d^{3}a}{dt^{3}}=-\frac{1}{H^{2}}\frac{d}{da}(aH^{2}q)=1-\frac{(1+z)}{\widetilde{H}^{2}}\frac{d\widetilde{H}^{2}
}{dz}
+\frac{(1+z)^{2}}{2\widetilde{H}^{2}}\frac{d^{2}\widetilde{H}^{2}
}{dz^{2}}
\end{equation}
\begin{equation}
S=\frac{1}{aH^{4}}\frac{d^{4}a}{dt^{4}}=\frac{1}{H^{3}}\frac{d}{da}(aH^{3}J)=-\frac{(1+z)^{2}}
{\widetilde{H}^{3}}\frac{d}{dz}\left(\frac{\widetilde{H}^{3}}{1+z}J\right)
\end{equation}
\begin{equation}
L=\frac{1}{aH^{5}}\frac{d^{5}a}{dt^{5}}=\frac{1}{H^{4}}\frac{d}{da}(aH^{4}S)=-\frac{(1+z)^{2}}
{\widetilde{H}^{4}}\frac{d}{dz}\left(\frac{\widetilde{H}^{4}}{1+z}S\right)
\end{equation}

which are usually referred to as the deceleration ($q$), jerk
($J$), snap ($S$) and lerk ($L$) parameters, respectively. The
Cosmographic Parameters's present day values (which are denote
with a subscript 0) may be used to characterize the evolutionary
status of the universe. For instance, $q_0 < 0$ denotes an
accelerated expansion, while $J_0$ allows to discriminate among
different accelerating models.

\begin{figure}
\includegraphics[scale=0.75]{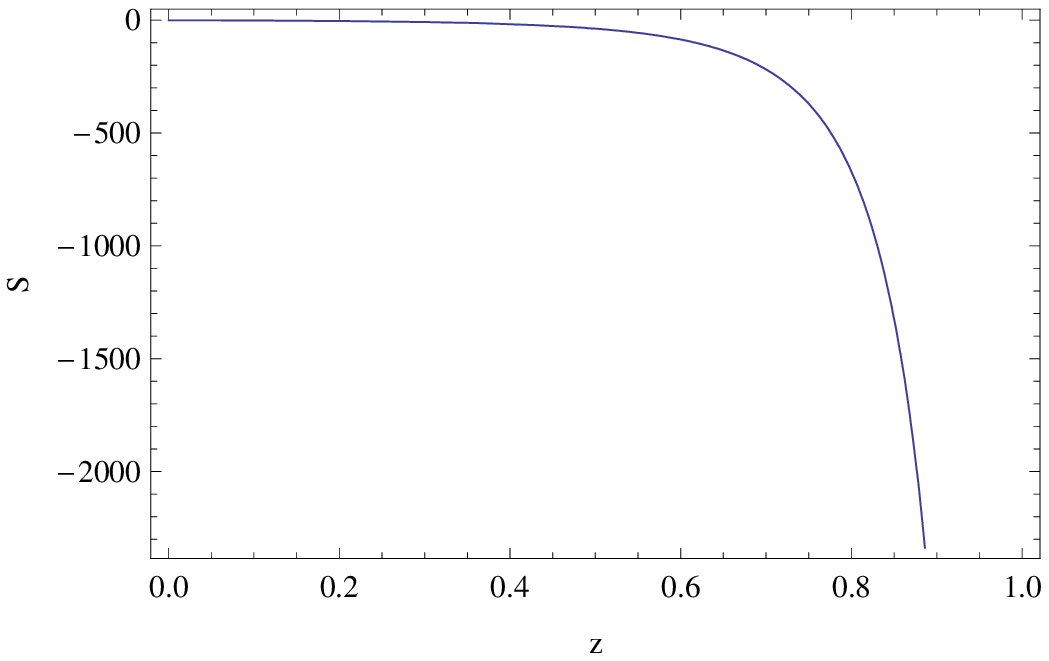}~~
\includegraphics[scale=0.75]{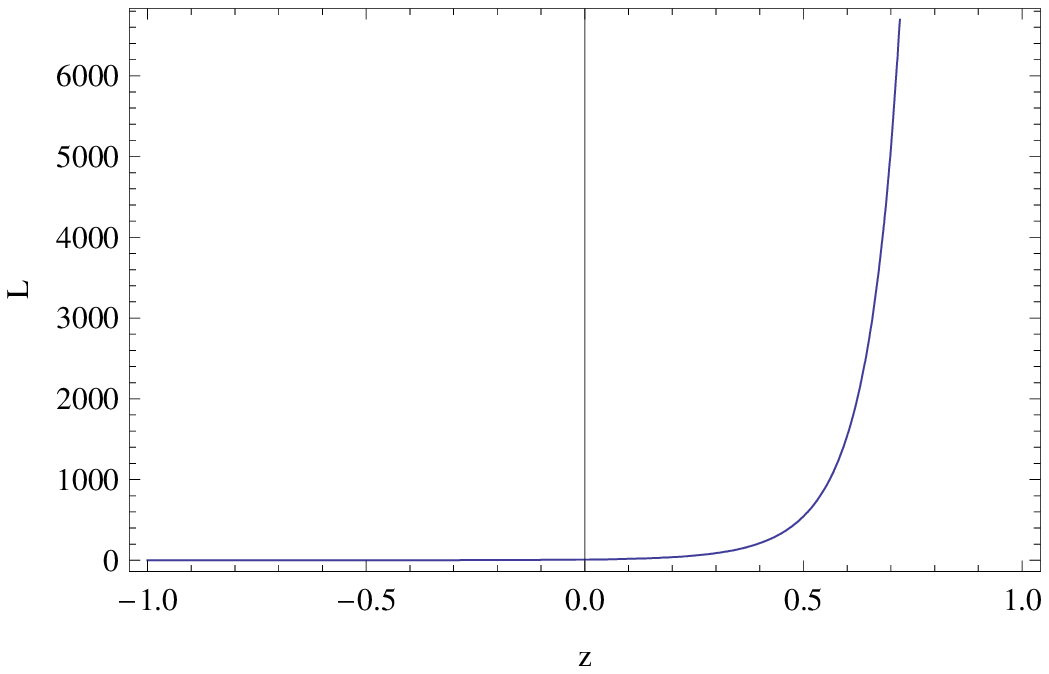}\\
\vspace{2mm}
~~~~~~~~~~~~~~~~~~~~~~~~~~~~~~~~~~~Fig.9~~~~~~~~~~~~~~~~~~~~~~~~~~~~~~~~~~~~~~~~~~~~~~~~~~~~~~~~~~~~~~Fig.10\\
\vspace{1mm} Fig. 9 shows the variation of the Snap parameter S
with redshift z. \\
Fig. 10 shows the variation of the Lerk
parameter L with redshift z.
\end{figure}

\subsection{$Om$ \normalsize\bf{Parameter}}

Recently, Sahni et al \cite{Sahni2003a,Sahni2008} proposed a new
cosmological parameter named {\it Om} which was introduced to
differentiate $\Lambda$CDM from other dark energy models  {\it Om}
diagnostic has been discussed together with statefinder for
generalized Chaplygin gas model from cosmic observations in
\cite{Lu2009,Tong2009}. Generally, it is a combination of Hubble
parameter and the cosmological redshift and provides a null test
of dark energy. For $\Lambda$CDM model, ${\it Om} = \Omega_{m0}$
is a constant, independent of redshift $z$. Also it helps to
distinguish the present matter density constant $\Omega_{m0}$ in
different models more effectively. The main utility for {\it Om}
diagnostic is that the quantity of {\it Om} can distinguish dark
energy models with less dependence on matter density
${\Omega_{m0}}$ relative to the EoS of dark energy. Our starting
point for {\it Om} diagnostics in the Hubble parameter and it is
defined as:

\begin{equation}
Om(z)=\frac{\widetilde{H}^{2}(z)-1}{(1+z)^{3}-1}
\end{equation}

Using the above result we have

\begin{equation}
Om(z)=\frac{\frac{\sqrt{-\frac{\beta}{a^{3(1+\omega_{m})}}}\left[\alpha
\sqrt{-\frac{a^{3(1+\omega_{m})}}{\beta}}
J_{-U}(V)C+\sqrt{-\frac{6a^{3\delta}\rho_{m0}}{\beta}}\left(2J_{-1+U}(V)+\left(J_{-1-U}(V)-
J_{1-U}(V)\right)C\right)\right]}{36H_{0}^{2}\left(J_{U}(V)+J_{-U}(V)C\right)}-1}{(1+z)^{3}-1}
\end{equation}

\begin{figure}
~~~~~~~~~~~~~~~~~\includegraphics[scale=0.85]{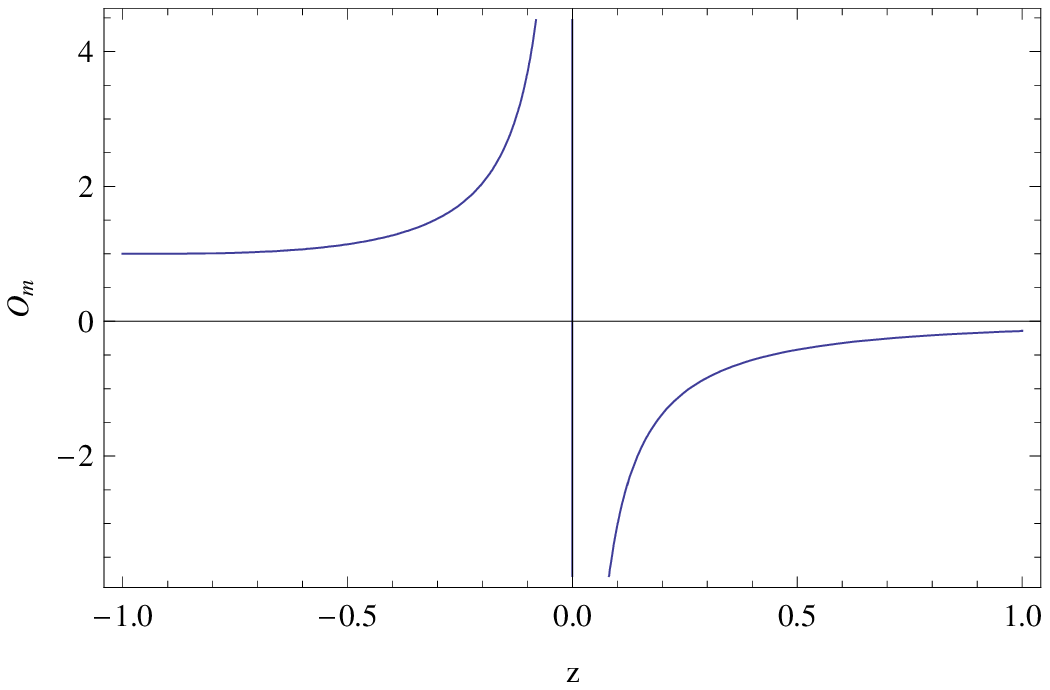}~~\\
\vspace{2mm}
~~~~~~~~~~~~~~~~~~~~~~~~~~~~~~~~~~~~~~~~~~~~~~~~~~~~~~Fig.11~~~~~~~~~~~~~~~~~~~~~~~~~~~~~~~~~~~~~~~~~~~~~~\\
\vspace{4mm}Fig.11 gives the plot of \textit{Om}-diagnostic
against redshift parameter $z$ \vspace{6mm}
\end{figure}

\section{\normalsize\bf{Stability Analysis}}

In this section, we have examined the stability of this model. In
this connection, the most important quantity is the squared speed
of sound which is denoted as $v_s^2$ and defined as a ratio of the
effective pressure and energy densities i.e.,
$v_s^2=\frac{\dot{p}}{\dot{\rho}}$. The sign of $v_s^2$ plays a
vital role for stability analysis of a background evolution of
cosmic models. It is well known that the model is stable if
$v_s^2>0$ and if $v_s^2<0$ implies that the model is classically
unstable\cite{Myung}. In 2008, Kim et al. \cite{Kim} found that
$v_s^2$ for agegraphic DE is always negative which leads to
classically instability of that model. Recently many researchers
are using this methods to analyzed the stability of that models on
which they worked and of which some authors
\cite{Jawad,Ebrahimi,Sharif,Setare} have reached to a conclusion
that HDE, ADE, NADE, Chaplygin gas, holographic Chaplygin,
holographic $f(T)$, holographic $f(G)$, new agegraphic $f(T)$, new
agegraphic $f(G)$ models are classically unstable because squared
speed of sound is negative i.e., $v_s^2<0$  throughout the
evolution of the universe. Here we consider
\begin{equation}
 v_s^2=\frac{\dot{p}_{MG}}{\dot{\rho}_{MG}}
\end{equation}
and plot $v_s^2$ versus $t$ by taking the power-law scale factor
in this model. Fig. 12 shows the variation of $v_s^2$ with $t$ for
the above mentioned cases.
\begin{figure}
~~~~~~~~~~~~~~~~~\includegraphics[scale=0.85]{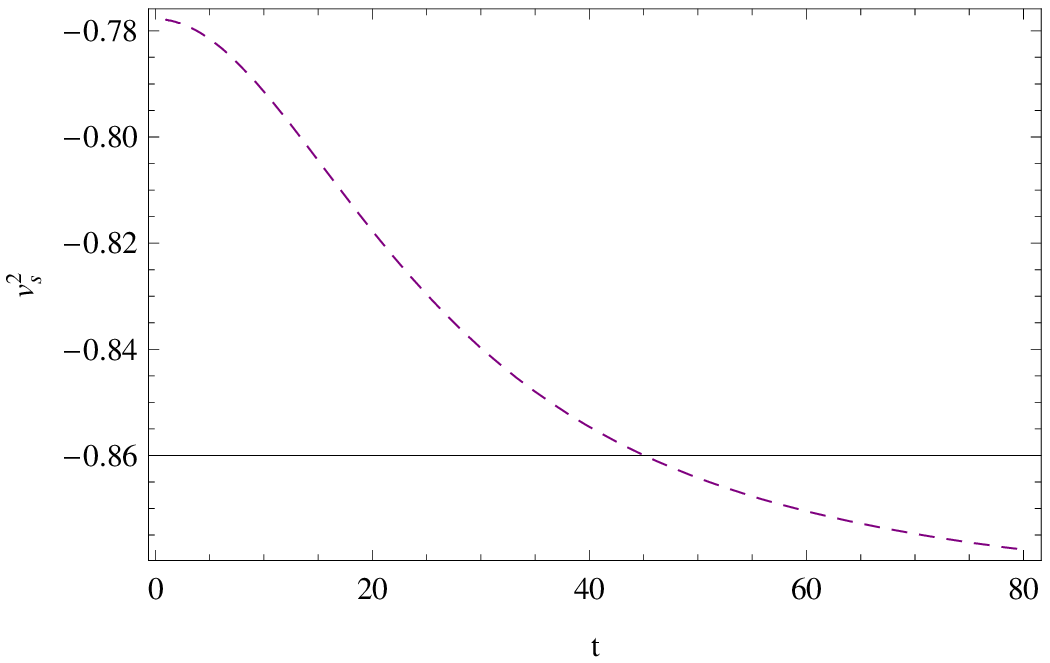}~~\\
\vspace{2mm}
~~~~~~~~~~~~~~~~~~~~~~~~~~~~~~~~~~~~~~~~~~~~~~~~~~~~~~Fig.12~~~~~~~~~~~~~~~~~~~~~~~~~~~~~~~~~~~~~~~~~~~~~~\\
\vspace{4mm} Fig 12 shows the variation of the squared speed of
sound $v_s^2$ against the cosmic time $t$.\vspace{6mm}
\end{figure}
In this model we observed that the square speed of sound remains
negative for the present and future epoch, which implies that our
predicted model for the above mentioned scenarios with power-law
form of scale factor is classically unstable.

\section{\normalsize\bf{Discussions}}

In this work we have studied the behaviour of some cosmological
parameters like Deceleration parameter, EoS parameter, {\it Om}
parameter, State-finder parameters and Cosmographic parameters in
an universe described by interacting new holographic dark energy
model in $f(T)$ gravity. We have considered a suitable interaction
between holographic dark energy and dark matter to account for the
coincidence scenario. We have derived the expressions for the
above mentioned cosmological parameters and plotted them against
the cosmological redshift parameter $z$.

In fig.1, we have plotted the deceleration parameter against the
redshift. From the plot we see that at the current epoch $(z=0)$,
the value of $q$ is around $-0.8$, which shows that the universe
is going through an accelerated expansion. From the curve it is
also evident that the universe is evolving from an early
decelerating phase to a late accelerating phase. In fig.2, the EoS
parameter is plotted against the redshift parameter. It is seen
that currently the value of $\omega$ is around $-0.9$, which lies
in the quintessence range $(-1<\omega<-1/3)$. Moreover it is seen
that near the $\omega=-1$ axis the curve assumes an asymptotic
behaviour, which shows that the universe never enters the phantom
regime, although the trajectory is infinitesimally close to the
phantom divide. In the figs. 3, 4 and 5, the trajectories in the
$\omega$--$ \omega'$ plane are obtained for different values of
interactions. It is seen that for no or smaller interaction models
we get freezing regions which are best suited for the current
accelerating scenario. But for high interactions we get thawing
models. In the figs. 6 and 7, the statefinder parameters $r$ and
$s$ are respectively plotted against the redshift parameter. From
the trajectories it is seen that in the present time $r=1, s=0$,
which corresponds to the standard $\Lambda$CDM cosmological model
of accelerating universe. In fig. 8, trajectories in the $r-s$
plane are obtained for the given model. This figure also confirms
the fact that when $r=1$, we have $s=0$, which is quite
characteristic for an accelerating model of the universe. In figs.
9 and 10, the Snap and the Lerk parameters are respectively
plotted against the redshift parameter. From fig.9, we see that
after passing through a series of negative values in the early
universe, it assumes constancy around the zero level in the
present universe. Almost similar is the case for the Lerk
parameter in fig 10, the only difference being its values lying in
the positive region in the early universe. In fig. 11, we have
obtained the trajectories for the $Om-$diagnostic. The
trajectories obtained are unique for the given model and can be
used to differentiate the model from others. Finally in fig. 12,
the squared speed of sound $v_s^{2}$ is plotted against cosmic
time $t$. From the figure it is seen that the trajectory for the
present model remains at the negative level thus exhibiting the
unstable nature of the model. Nevertheless the study reveals that
the model is perfectly consistent with the notion of the
accelerating universe, at least for low interaction cases, and the
derived values of all the parameters comply to this fact.\\

{\bf Acknowledgement:}\\

The authors are thankful to IUCAA, Pune, India for warm
hospitality where a part of the work was carried out.\\

\end{document}